\documentclass{article}

     \PassOptionsToPackage{numbers, compress}{natbib}
     \usepackage[final]{neurips_2019}

\usepackage[utf8]{inputenc} 
\usepackage[T1]{fontenc}    
\usepackage{hyperref}       
\usepackage{url}            
\usepackage{booktabs}       
\usepackage{amsfonts}       
\usepackage{nicefrac}       
\usepackage{microtype}      
\usepackage{subfigure}      
\usepackage{caption}
\usepackage{graphicx}
\usepackage{amsmath}
\usepackage{booktabs, tabularx}
\usepackage{float}
\usepackage{multicol}

\title{Effective Sub-clonal Cancer Representation to Predict Tumor Evolution}

\author{
  Adnan Akbar$^1$, Geoffroy Dubourg-Felonneau$^1$, Andrey Solovyev$^{1,2}$, \\
  \textbf{John W Cassidy$^1$, Nirmesh Patel$^1$, Harry W Clifford$^1$} \\
  \\
  \and
  $^1$Cambridge Cancer Genomics \\
  Cambridge, UK \\
  www.ccg.ai
  \and
  $^2$The University of Edinburgh \\
  Edinburgh, UK \\
}

\begin{document}

\maketitle

\begin{abstract}

The majority of cancer treatments end in failure due to Intra-Tumor Heterogeneity (ITH). ITH in cancer is represented by clonal evolution where different sub-clones compete with each other for resources under conditions of Darwinian natural selection. Predicting the growth of these sub-clones within a tumour is among the key challenges of modern cancer research. Predicting tumor behavior enables the creation of risk profiles for patients and the optimisation of their treatment by therapeutically targeting sub-clones more likely to grow. Current research efforts in this space are focused on mathematical modelling of population genetics to quantify the selective advantage of sub-clones, thus enabling predictions of which sub-clones are more likely to grow. These tumor evolution models are based on assumptions which are not valid for real-world tumor micro-environment. Furthermore, these models are often fit on a single instance of a tumor, and hence prediction models cannot be validated. 

\smallskip

This paper presents an alternative approach for predicting cancer evolution using a data-driven machine learning method. Our proposed method is based on the intuition that if we can capture the true characteristics of sub-clones within a tumor and represent it in the form of features, a sophisticated machine learning algorithm can be trained to predict its behavior. The steps implemented in our approach are:

\begin{itemize}
  \setlength{\itemsep}{3pt}%
  \item Longitudinal data acquisition - labelled longitudinal data, with known outputs for validations, from the TRACERx lung cancer liquid biopsy study to apply machine learning and predict cancer evolution.
  \item Sub-clonal profiling - Inference of the number of sub-clones, and the number of mutations within each sub-clone, for each tumor.
  \item Feature selection - Features which can truly represent sub-clones characteristics were identified. Based on mutation location we extracted several features including functional impact, properties of genes where mutations occur, and the properties of associated proteins and their interactions with other proteins.
  \item Feature representation as a fixed-length vector - Different sub-clones have different number of mutations and feature representation at a mutation level needed to be aggregated at a sub-clone level to be used by a model.
  \item Machine learning model - Both regression analysis (to predict the amount of growth or reduction of each sub-clone in future) and classification analysis (to predict if a sub-clone will grow or not in future) were applied. With regression analysis we obtained an $R^2$ of 0.14 and with classification analysis we obtained an $AUC-ROC$ of >0.6. Both were better than mean and random predictions respectively showing predictive strength in cancer evolution.
\end{itemize}

The work presented here provides a novel approach to predicting cancer evolution, utilizing a data-driver approach. We strongly believe that the accumulation of data from microbiologists, oncologists and machine learning researchers could be used to encapsulate the true essence of tumor sub-clones, and can play a vital role in selecting the best cancer treatments for patients.

\end{abstract}

\section{Introduction}
Cancer is an evolutionary process where cellular sub-populations known as sub-clones compete with each other under conditions of Darwinian natural selection\cite{yates2015subclonal}. Resultant Intra-Tumor Heterogeneity (ITH) has been associated with higher likelihood of relapse and increased resistance to therapy in cancer patients\cite{greaves2012clonal, o2015imaging}. The ability to precisely predict how these sub-clones will evolve over time can help clinicians to develop an effective cancer treatment and reduce treatment failures\cite{chkhaidze2019spatially}.

Tumor evolution in literature has mostly been addressed through the mathematical modelling approach. One early analytical approach is based on using models from population genetics to quantify the selective advantage of sub-clones, thus enabling predictions of which sub-clones are more likely to grow \cite{nielsen2003estimating}. Until recently, it was almost impossible to observe the clonal growth at different time points due to the invasive nature of tissue biopsies. Therefore, the analysis of tumour growth and mutation frequencies was normally conducted at one time point - from the excised tumour sample.

Next Generation Sequencing (NGS) methods such as targeted sequencing or full exome sequencing, followed by variant calling algorithms, allow the calculation of Variant Allele Frequencies (VAFs) in a sample. A number of quantitative models have been developed to infer the evolutionary events that generated the resulting distribution of allele frequencies \cite{williams2018quantification, graham2017measuring}. These methods allow us to determine whether there is selection occurring in the tumour by analysing the VAF spectra.

All of these tumor evolution models are validated either by fitting the mathematical model on a single instance of tumor VAFs or by using simulated material. The validations of these models were not done on real-world longitudinal samples.

In this paper, we present a novel data-driven approach to predict cancer evolution for real-world data. Our proposed method is based on the intuition that if we can capture the true characteristics of sub-clones within a tumor and represent it in the form of features, a sophisticated machine learning algorithm can be trained to predict its behavior. Our main challenges are as follows:

    \begin{itemize}
        \setlength\itemsep{0.5em}
        \item Accurately identifying the \textbf{number of sub-clones} and their formation and constitution in a given tumor.
        \item Feature-based \textbf{representation of sub-clones} to encapsulate the characteristics of underlying mutations towards tumorigenesis.
        \item Gathering \textbf{labelled data across timepoints} for application of machine learning models to predict evolution.
    \end{itemize}

\section{Methods}

    \begin{figure}[H]
          \centering
            \includegraphics[width=0.95\linewidth]{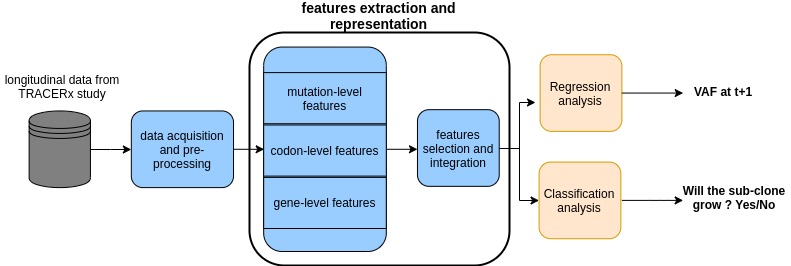}\\
            \caption{An overview of the proposed method}
    \end{figure}
    
\subsection{Data acquisition and pre-processing}

We used longitudinal data from the TRACERx lung cancer liquid biopsy study by Abbosh et al. (2017).\cite{Abbosh2017PhylogeneticCA}. This includes data from 24 patients monitored for 1-3 years. For these patients, DNA mutational data was generated by liquid biopsy (2-8 samples per patient) and subclonal information was generated with PyClone\cite{roth2014pyclone}.

\subsection{Feature extraction}

\begin{figure}[H]
          \centering
            \includegraphics[width=.85\linewidth]{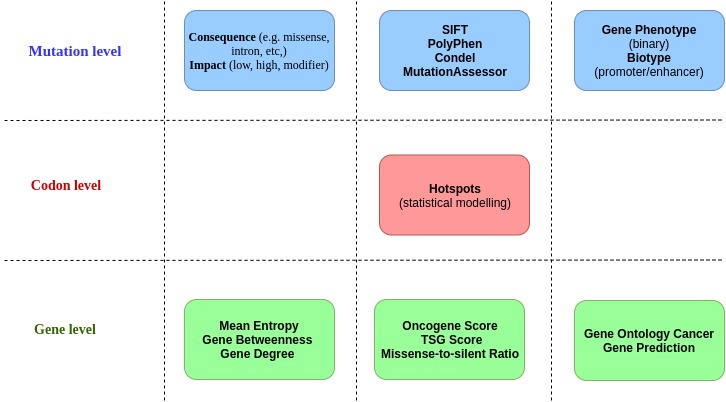}
            \caption{Final features extracted}
            \label{fig:features}
    \end{figure}

Features were extracted at three levels: mutation, codon, and gene (Figure \ref{fig:features}).

For mutation-level features we utilized the Ensembl Variant Effect Predictor\cite{McLaren2016TheEV}, which provides individual scores around likely consequences and potential impacts for a given mutation, including scores predicting protein effect such as SIFT and PolyPhen.

For codon-level features we applied a statistical model to data from large-scale genomics projects, The Cancer Genome Atlas \cite{Collins2007TheCG} and Genie \cite{aacr2017aacr}, to give a hotspot score per codon based on frequency of observed mutations against background mutability.

For gene-level features we utilized the various scores and metrics produced by the 20/20+ tool \cite{Tokheim2016EvaluatingTE}, which calculates these as part of a machine-learning-based ratiometric method of driver-gene classification through evaluation of the proportion of inactivating mutations and recurrent missense mutations in a gene of interest.
    
\subsection{Feature selection and integration}
As different sub-clones have differing numbers of mutations, statistical features including mean, max, min and median were calculated from the sets of mutations to give fixed length feature vectors to represent the sub-clones. The final form of the features matrix is as follows:

\begin{multicols}{2}
    \[features\_matrix=\begin{bmatrix}
        x_{11} & \dots  & x_{1f} \\
        \vdots & \ddots & \vdots \\
        x_{c1} & \dots  & x_{cf}
    \end{bmatrix}\]
    \\
    $c$ : number of sub-clones\\
    $f$ : number of features\\
\end{multicols}

\subsection{Model training and evaluation}
Clonal-evolution within a tumor is predicted using two approaches: a regression analysis and a classification analysis. We evaluate both approaches with a range of algorithms, hyperparameter optimization, and K-fold cross validation at k=5.

\section{Results}

\subsection{Regression Analysis}

\begin{figure}[H]
        \centering
        \centering\includegraphics[width=.49\linewidth]{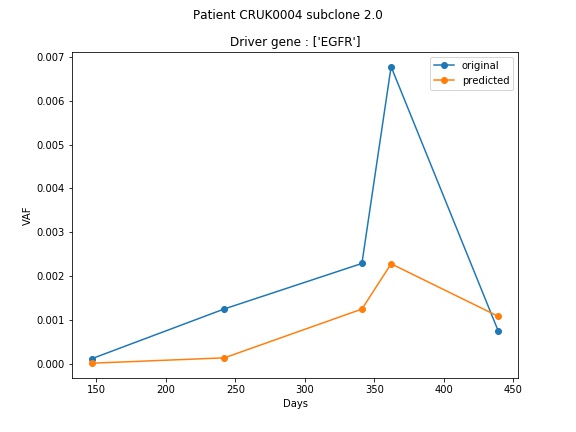}
        \centering\includegraphics[width=.49\linewidth]{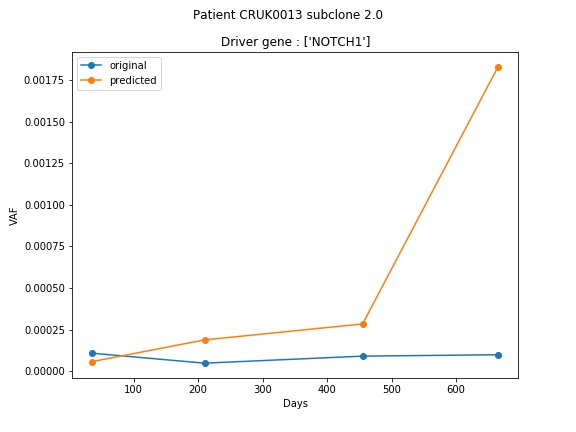}
        \centering\includegraphics[width=.49\linewidth]{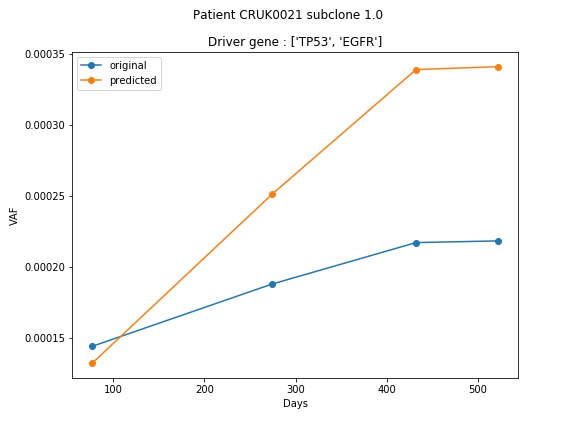}
        \centering\includegraphics[width=.49\linewidth]{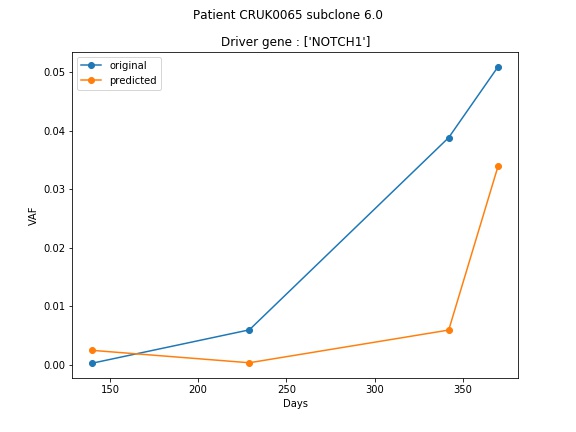}
        \caption{Predicting change in VAF over time (Days) for four randomly chosen sub-clones}
        \label{fig:regression}
    \end{figure}

A random forest regression analysis returned promising results (Figure \ref{fig:regression}). Overall, an $R^2$ of 0.16 was achieved. Although there is much to be improved in terms of model performance, this shows promising signal, indicating a better than random performance.

\subsection{Classification Analysis}

As a classification problem the approach is simplified greatly, with the goal of simply predicting if the sub-clonal VAF will increase or decrease. Classification methods used were logistic regression, random forest, support vector machine (SVM) and artificial neural network, respectively. We present results for predicting change of sub-clones, PCA of sub-clones, and driver mutations, respectively (Table \ref{tab:classification}).

\bgroup
\def\arraystretch{1.5}
    \begin{table}[H]
        \centering
        \caption{Performance summary of classification models (ROC-AUC; F1 score)\medskip}
        \label{tab:classification}
        \begin{tabular}{|c|c|c|c|c|}
            \hline
                & \textbf{\begin{tabular}[c]{@{}c@{}}Logistic Regression\end{tabular}} & \textbf{\begin{tabular}[c]{@{}c@{}}Random Forest\end{tabular}} & \textbf{\begin{tabular}[c]{@{}c@{}}SVM\end{tabular}} & \textbf{\begin{tabular}[c]{@{}c@{}}ANN\end{tabular}} \\
            \hline
                Sub-clones & 0.298; 0.399 & 0.354; 0.333 & 0.422; 0.389 & 0.512; 0.623 \\
            \hline
                PCA sub-clones & 0.456; 0.519 & 0.573; 0.527 & 0.511; 0.476 & 0.505; 0.421 \\
            \hline
                Driver mutations & 0.618; 0.348 & \textbf{0.785; 0.643} & 0.566; 0.488 & 0.511; 0.364 \\
            \hline
        \end{tabular}
    \end{table}
\egroup

\section{Conclusions \& Future Work}

In this paper, we have shown that is possible to predict the course of tumor evolution using prior information. This has profound implications in the clinical management of cancer as a chronic disease. Interestingly, our models performed best on 'driver' mutations, i.e. those thought to be primarily responsible for driving clonal growth. This adds weight to the biological significance of the results. 

Clearly there is much to be improved in terms of model performance. Current and future work by our group will include greater data collection, applying improved liquid biopsy analytical methods developed by Cambridge Cancer Genomics, and further refinement of the evolutionary model choice, which was found to greatly influence performance. Overall, this enable more accurate prediction of cancer evolution through data-driven machine learning approaches, and ultimately greatly improve cancer care. 

\bigskip
\bibliography{bib}

\end{document}